# Field-free high-frequency exchange-spring spin-torque nano-oscillators

Sheng Jiang,[1,2,3,4,*] Sunjae Chung,[5] Quang Tuan Le,[2] Ping Kwan Johnny Wong,[3] Wen Zhang,[3,*] and Johan Åkerman[2,6,7,*]

1. School of Microelectronics, South China University of Technology, 510641 Guangzhou, China

2. Department of Physics, University of Gothenburg, 41296 Gothenburg, Sweden

3. School of Microelectronics, Northwestern Polytechnical University, 710072 Xi'an, China

4. Yangtze River Delta Research Institute of NPU, Taicang, 215400 Jiangsu, China

5. Department of Physics Education, Korea National University of Education, 28173 Cheongju, Korea

6. School of Engineering Sciences, KTH Royal Institute of Technology, 100 44 Stockholm, Sweden

7. NanOsc AB, 16440 Kista, Sweden

E-mail: jiangsheng@scut.edu.cn; zhang.wen@nwpu.edu.cn; johanakerman@gu.se

# Abstract

Spin-torque nano-oscillators (STNOs) are a type of nanoscale microwave auto-oscillators utilizing spin-torque to generate magnetodynamics with great promise for applications in microwaves, magnetic memory, and neuromorphic computing. Here, we report the first demonstration of exchange-spring STNOs, with an exchange-spring ([Co/Pd]-Co) reference layer and a perpendicular ([Co/Ni]) free layer. This magnetic configuration results in high-frequency (>10 GHz) microwave emission at a zero magnetic field and exchange-spring dynamics in the reference layer and the observation of magnetic droplet solitons in the free layer at different current polarities. Our demonstration of bipolar and field-free exchange-spring-based STNOs operating over a 20 GHz frequency range greatly extends the design freedom and functionality of the current STNO technology for energy-efficient high-frequency spintronic and neuromorphic applications.

Spin-torque nano-oscillators (STNOs) [1,2] are spintronic nano-devices utilizing spin torques to excite spin precession. These nanoscale nonlinear oscillators are capable of generating various magnetodynamic modes, such as propagating spin-waves [3–5], localized bullets, [5,6] magnetic droplets, [7–12] dynamic vortex [13,14] and dynamic skyrmions. [15] During the last decades, tremendous attention has been paid to their applications in, for instance, signal generators, [1] spectrum analyzers, [16,17] spin diodes, [18–20] wireless communication, [21–23] microwave-assisted magnetic random-access memories, [24,25] Ising machines [26] and neuromorphic computing, [27–29] due to their capability of generating highly-tunable ultrawide microwave signals ranging from hundreds of megahertz to tens of gigahertz. [30]

Generally, an STNO consists of a nonmagnetic spacer sandwiched between two ferromagnetic layers, with one as a free layer and the other as a reference layer. Magnetic anisotropies of those ferromagnetic layers are key in determining the operation performances in STNOs. For instance, in STNOs with in-plane magnetic anisotropic (IMA) ferromagnetic layers, the microwave linewidth and power can be improved dramatically by synchronizing the propagating spin waves. [31,32] Nevertheless, a high magnetic field is generally required in these in-plane STNOs to produce a non-collinear magnetic configuration, for the sake of maximizing the current-induced spin torques and detecting the auto-oscillating signals. [5,33] Using materials with perpendicular magnetic anisotropy (PMA) as the free layerand/or the reference layer [34–36] may be a solution to this issue, where novel dynamic mode---magnetic droplet---can even be generated at a low field regime. [37,38]

Besides IMA and PMA, STNOs with tilted magnetization have been proposed theoretically, promising a high tunability in microwave frequencies at low-to-zero field regime. [39–44] In practice, such tilted magnetization can be realized by coupling IMA and PMA materials through an interface exchange interaction, *i.e.*, the so-called exchange spring (ES). [45–54] Various magnetic easy axes can then be obtained, based on a tailored balance between the strength of IMA, exchange coupling, and the PMA.

Here, we report the first demonstration of exchange-spring STNOs with a carefully optimized structure, [Co/Ni]/Cu/[Co/Pd]-Co. The [Co/Pd]-Co exchange spring is composed of a [Co/Pd] multilayer hard magnet with strong PMA and a soft Co layer with easy IMA, while the free layer consists of a PMA [Co/Ni] multilayer. By systematically analyzing the dynamical behavior at various amplitudes and angles of external magnetic fields and *dc* currents, we find that the STNO can produce high frequency microwave signals over 20 GHz and the frequency can still reach up to 10.5 GHz in the absence of magnetic fields in the exchange spring layer. The precession frequency of magnetic droplet solitons can be observed in the [Co/Ni] free layer. Moreover, this type of oscillator can be operated at both negative and positive injecting currents by selectively activate the [Co/Ni] or [Co/Pd]-Co layer, thus holding a potential of serving as a bipolar oscillator.

## RESULTS AND DISCUSSIONS

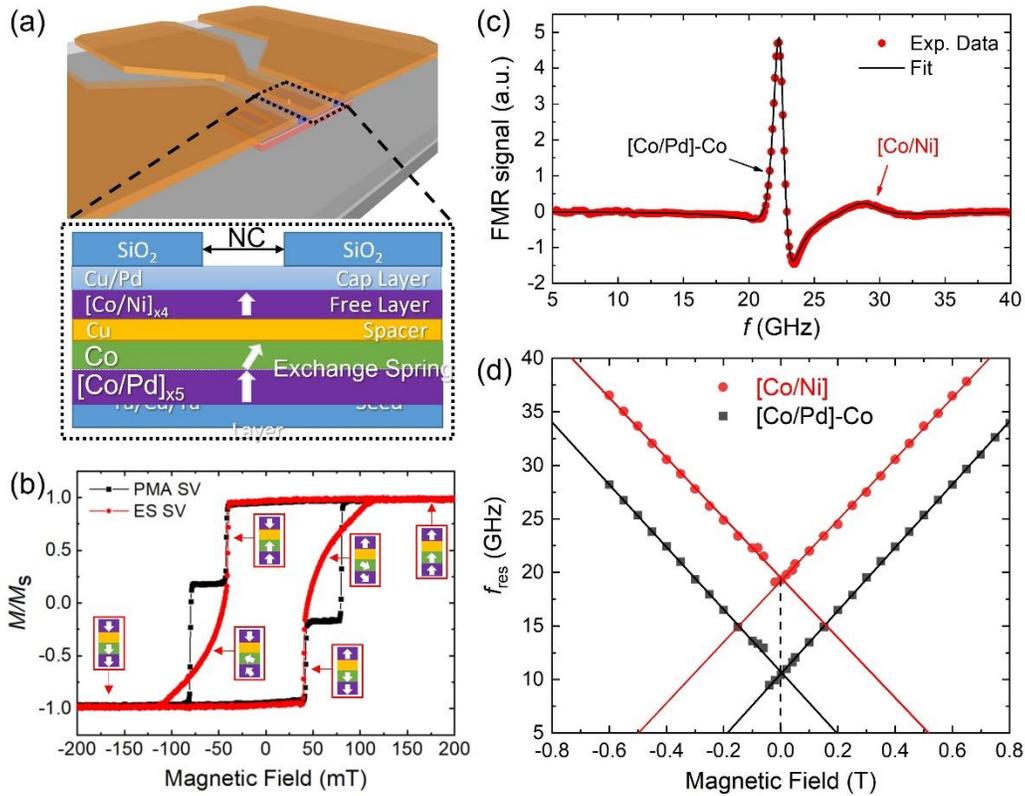

**Figure 1** (a) Schematic picture of a nano-contact spin-torque nano-oscillator (NC-STNO) device and its film stack structure, where the exchange-spring layer consists of a hard magnet [Co/Pd]x5 with strong PMA and a soft Co layer with easy IMA; (b) Hysteresis loops of the exchange spring (ES) ([Co/Pd]-Co/Cu/[Co/Ni], red dot-curve) and all-PMA ([Co/Pd]/Cu/[Co/Ni], black square-curve) spin valves (SVs); (c) Typical FMR spectrum measured by sweeping the frequency at a perpendicular magnetic field of 0.4 T. The spectrum is fitted with Lorentz functions shown as the black solid line. (d) The resonance frequencies of the two peaks vs. magnetic field and fits using the Kittel equation (solid lines).[60]

**Figure 1**a shows a schematic picture of the nanocontact (NC) STNO with a GSG top electrode (light brown). The initial spin valve comprises an exchange spring reference layer with [Co/Pd]-Co, a Cu spacer, and a PMA free layer with [Co/Ni]. The detailed stack information can be seen in **Methods**. The spin valve was then patterned into a rectangular shape of 16×8 µm$^2$ with the nanocontact positioned at the center of the stack. **Figure 1**b shows hysteresis loops measured

by the alternating grating magnetometer for the exchange-spring spin valve (red curve) and all-PMA spin valve (black curve) consisting of [Co/Pd]/Cu/[Co/Ni] for comparison. Two-step switching is observed for the all-PMA spin valve, corresponding to the switching fields of the [Co/Ni] free layer at 42 mT and the [Co/Pd] reference layer at 80 mT, respectively. To form an exchange-spring reference layer, a 3-nm Co layer was added on top of the reference layer [Co/Pd] as in **Figure 1**a. Its hysteresis loop, represented by the red curve in **Figure 1**b, exhibits a steep change of magnetic moment at about 42 mT, as the switching of the [Co/Ni] free layer remains essentially unaffected by the addition of the Co layer. Then, the magnetic moment shows a smooth continuous increase with the magnetic field, suggesting a gradual change of the magnetization of the exchange-spring layer instead of sharp switching. The magnetic configurations at different magnetic fields are presented as the schematics in Figure 1b. The magnetic anisotropy of the exchange spring with [Co/Pd]-Co has been studied systematically in Ref. [50]. The exchange-spring layer here was deposited with the same thickness and sputtering conditions with Ref. [50] to achieve a tilted magnetic anisotropy. The hysteresis loop confirms that a reference layer exchange spring is achieved.

Then, we conduct FMR measurements of the exchange-spring spin valve by sweeping the frequency at constant magnetic field, to avoid any field-induced spin direction changes to the exchange-spring layer. A typical spectrum measured at a fixed perpendicular magnetic field of 0.4 T is illustrated in **Figure 1**c, from which two characteristic peaks are observed. The strong peak at 22.3 GHz originates from the exchange spring, while the less intense peak at 30.6 GHz corresponds to the [Co/Ni] free layer. These peaks are different in linewidth, due to the smaller damping ($\alpha_{eff}^{ES} = 0.0171$, see detailed calculation in **Supporting Information**) of the exchange spring than that of [Co/Ni] ($\alpha_{eff}^{[Co/Ni]} = 0.0455$), [55] and thus a narrower linewidth of the strong peak. Similarly, the inhomogeneous linewidth of the [Co/Pd]-Co exchange-spring layer ($\Delta H_0^{ES} = 40$ mT) is about one third of the [Co/Ni] ($\Delta H_0^{[Co/Ni]} = 120$ mT) [55], making its associated peak narrower. By fitting the FMR spectra at different fixed magnetic fields, two resonance peaks have been extracted and plotted in **Figure 1**d. We then fit $f_{res}$ with the out-of-plane Kittel equation, [56]

$$f_{res} = \frac{\gamma \mu_0}{2\pi}(H - M_{eff}) \tag{1}$$

where $\mu_0$ is the permeability of free space, $\frac{\gamma}{2\pi}$ is the gyromagnetic ratio, and $M_{eff}$ is the effective magnetization. The effective magnetization $\mu_0 M_{eff} = -0.66$ T for the broad and higher frequency peaks is in excellent agreement with our previous FMR study on the PMA spin valve with the same structures of [Co/Ni] free layer. [55,57] The resonance frequencies ($f_{res}$) from the narrower and lower frequency peak (black dots) have also been fitted with **Equation** 1, yielding a value of $\mu_0 M_{eff} = -0.35$ T for [Co/Pd]-Co reference layer.

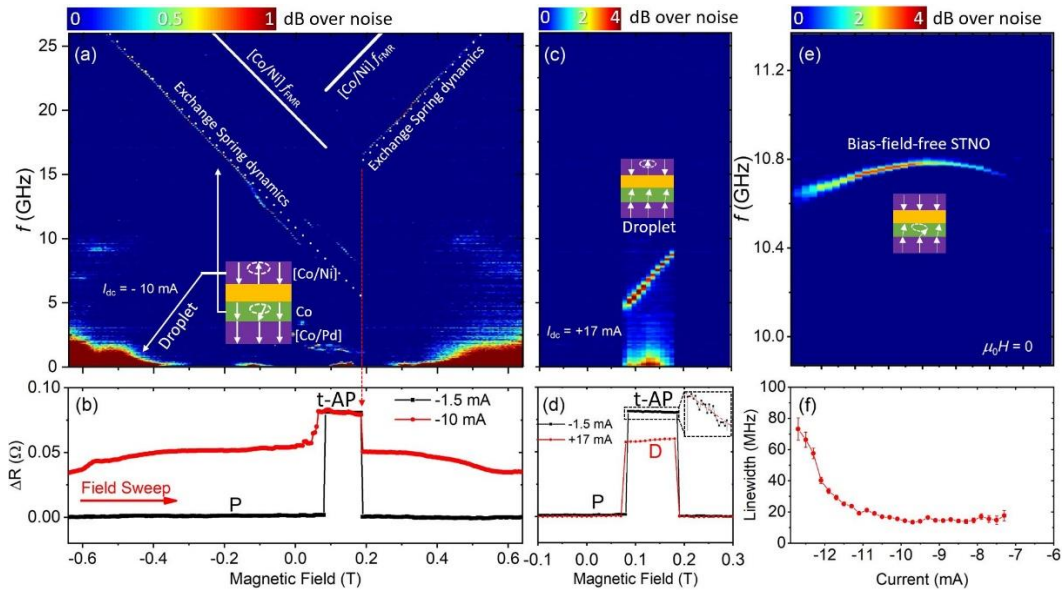

**Figure 2** (a) Power Spectrum Density (PSD) as a function of the magnetic field measured at -10 mA. The white solid and dashed lines indicate the calculated FMR frequency of the [Co/Ni] free layer and the [Co/Pd]-Co exchange-spring layer from measurements in Fig.1(d). The inset is a schematic of the magnetodynamics of the two layers. (b) The corresponding magnetoresistance (MR) as a function of the magnetic field measured at -1.5 and -10 mA. (c, d) PSD and related MR as functions of the magnetic field at +17 mA. The magnetic fields were normal to the film plane ($\theta_H = 90°$). (e, f) PSD and linewidth as functions of current at a zero field, respectively. P, t-AP and D in (b) and (d) stand for parallel, tilted-antiparallel and droplet states, respectively.

Having confirmed the properties of our exchange-spring spin valve, we now turn to the auto-oscillating properties of a NC-STNO device with NC radius $R_{NC} = 60$ nm. **Figure 2** shows the power spectrum density (PSD) and magnetoresistance (MR) as functions of perpendicular magnetic field ($\theta_H = 90°$). At low current ($I_{dc} = -1.5$ mA), the spin transfer torque (STT) is too small to trigger any magnetodynamic behavior. Therefore, the device is in a parallel (P) state with low MR. Sweeping magnetic field from 0.65 to 0.08 T results in the switching of the free layer [Co/Ni] and thus the tilted-antiparallel (t-AP) state with high MR, as marked by the black curve in **Figures 2**b and d. That this state is indeed t-AP and not fully AP can be seen from the linear slope of MR, *i.e.* the resistance decreases linearly with increasing field as the tilt angle of the Co layer increases away from the t-AP state in the insert of **Figure 2**d.

At stronger negative current ($I_{dc} = -10$ mA; red curve), the P state is replaced by an intermediate resistance state (the t-AP state remains unaffected), indicative of a partially reversed core in the free layer due to the STT, consistent with the nucleation of a free layer magnetic droplet soliton ("droplet", from hereon) observed in all-PMA STNOs. [10] This explanation is directly corroborated by the intense low-frequency microwave noise appearing at high fields, which is strong evidence of an unstable droplet. [10] [34]

However, in contrast to the all-PMA STNOs, another high-frequency signal shows up with a linear magnetic field dependence. By plotting the fitted exchange-spring FMR frequency from **Figure 1**d together with the PSD in **Figure 2**a, the $f_{FMR}^{ES}$ of the exchange-spring layer (the white dots in **Figure 2**a) is very close to the auto-oscillating frequency and exhibits an identical field dependence. The signal also jumps to a higher frequency at 0.19 T, *i. e.* at exactly the switching field of the exchange-spring layer. All these observations strongly suggest that this signal originates from the dynamics of the exchange-spring layer. In addition, the dynamics of the exchange-spring layer is only observed when the magnetization of the free layer is partially reversed (droplet) or fully reversed (t-AP), *i.e.* when there is a spin polarization consistent with destabilizing the reference layer.

We now turn to the dynamics in the t-AP state, where a positive current should excite the free layer, whereas a negative current should excite the reference layer. In **Figure 2**c we apply a high positive current ($I_{dc} = +17$ mA) and make several observations consistent with a droplet. First, the resistance is again at an intermediate value, as expected for a droplet. It is noteworthy, that the *slope* of the resistance vs. field has now changed sign compared to the t-AP state, consistent with the reversed magnetization of the droplet. In contrast to the droplet forming in the high-field P state (**Figure 2**a), where the all-perpendicular symmetry of the state cancels out the droplet microwave signal [10,34,35], here, the microwave signal from the droplet is directly observed, again confirming the low-field tilted state of the Co layer with an in-plane component of its magnetization, where it is also more affected by the Oersted field. [11,58] Finally, the droplet again comes with low-frequency microwave noise, as it moves around in the NC region.

Apart from the field-dependent droplet dynamics at positive currents, we have found that a high-frequency signal can be also observed at negative current in low-to-zero fields (**Figures 2**e-f). By applying a negative current at zero field, we achieve a clear microwave signal at around 10.8 GHz, much higher (1.8 times) than that of any other field-free STNOs (~6 GHz). [59] The observed linewidth is as narrow as 18 MHz in **Figure 2**f, again far better than *e.g.* field-free STNOs based on magnetic tunnel junctions. The exchange-spring STNO is hence promising for entirely field-free applications of energy-efficient microwave generators and neuromorphic device based on STNO.

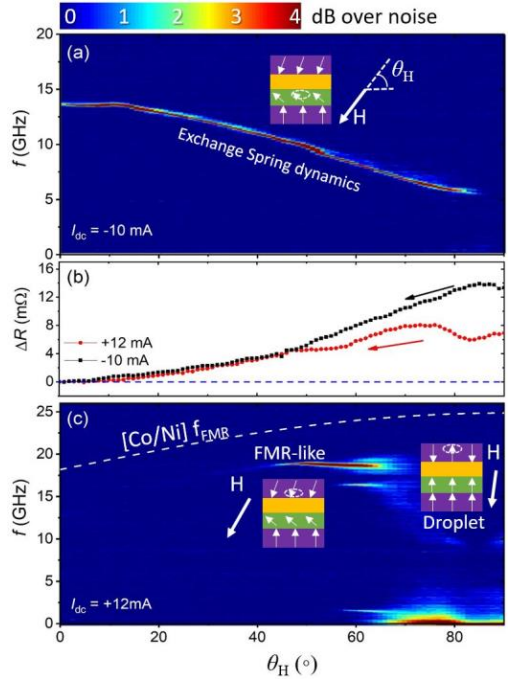

**Figure 3** (a) and (c) PSD as functions of angles of the magnetic field at -10 mA and +12 mA, respectively. (b) The corresponding MR of (a) and (c). The angle of the applied field is swept from $\theta_H = 90°$ to $0°$. The initial magnetic configuration was set to the t-AP state with the free layer magnetization aligned to the negative field direction. The insets are the schematics of magnetodynamics.

As described above, achieving a noncollinear configuration between the magnetization of the reference layer and the free layer is crucial for detecting the auto-oscillating signal of the NC-STNO [3]. It is therefore interesting to continue to examine the angular dependence of the observed microwave signals. Inspired by the results of **Figure 2** that the signals are more likely to be seen at partially/fully reversed states of the free layer, we first initialize the magnetic configuration to the t-AP state by applying a strong field ($\mu_0 H = 0.6$ T) and then ramping down to -0.15 T at $\theta_H = 90°$. **Figure 3** shows the PSD and MR results at $\mu_0 H = 0.19$ T, from which several features can be observed. **Figure 3**a are presented as line profiles in **Supporting Information**. In **Figures 3**a and b, clear signals exist at all the measured angles, with the frequency decreasing with increasing angle. This can be understood by the fact that the applied magnetic field is opposite to the magnetization direction of the reference layer and hence competes with the exchange coupling field (see the inset of **Figure 3**a). Another feature is the increasing MR with angles due to the increase of relative angle between the free and reference layers as the angle of magnetic field increases. When the positive current is applied at this t-AP state in **Figure 3**c, it first shows a weak signal at 10 GHz and a strong low-frequency noise at 90°, and the MR (the red curve in **Figure 3**b) is much lower than that at $I_{dc} = +17$ mA (the black curve in **Figure 3**b), which can be defined as a droplet mode. As the angle decreases to 70°, the low-f noise increases because of the strong drift instability of magnetic droplets in tilted fields. [9] The droplet modes gradually transit to higher frequency at around 70°. Then, it turns to fully high frequencies, where two peaks emerge at angles between 57° and 70°. This is due to the normal FMR-like mode, which could either be localized bullets or propagating spin waves. [5] The two peaks observed could be related to 1) hopping between two closely spaced auto-oscillation modes, as the sputtered [Co/Ni] films are polycrystalline, grain boundaries underneath the NC can impact the auto-oscillation modes from different locations; [60] and/or 2) the current-induced Oersted field results in an asymmetric energy landscape that can host the different spin-wave modes at varying locations. The frequency associated with these different spin-wave modes is likely non-identical, similar to the case of the NiFe free layer, where both modes can coexist. [5]

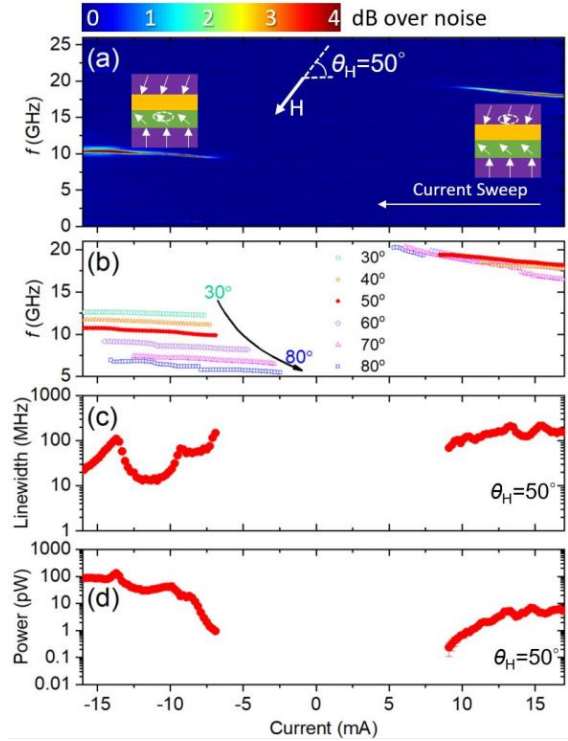

**Figure 4.** PSD as functions of dc current swept from $I_{dc}$ = +17 mA to -16 mA measured at $\theta_H = 50°$ and $\mu_0 H = 0.19\ T$. The insets are the schematics of magnetodynamics. (b) Extracted frequencies at different angles from $\theta_H = 30°$ to 80°. (c, d) Extracted linewidth and power as functions of applied currents at $\theta_H = 50°$. The spectra were fitted with Lorentz functions, and the power is calculated by taking the impedance mismatch and 40 dB pre-amplification into consideration.

**Figures 2** and **3** show the auto-oscillating signals at both positive and negative currents at different field amplitudes and angles. This feature raises the possibility of using the exchange-spring STNOs as a bipolar oscillator, since it can operate at both negative and positive currents. Here we set the amplitude of the magnetic field at -0.19 T and the angle at 30-80° to achieve the bipolar control of STNOs. Starting at the t-AP state, with a strong positive current, a high-frequency signal (17—20 GHz) is present, corresponding to the [Co/Ni] free layer dynamics. On the other hand, as the current decreases, the signal gradually disappeared at +8.4 mA in **Figure 4**a. Sweeping to negative current, another signal at about 10 GHz shows up from -6.2 mA till -16.2 mA. This signal results from the exchange-spring layer, whose frequency is identical to the frequency in

**Figure 3**a. We also extract the frequencies at different angles (30°—80°) by fitting the spectra with the Lorentz function in **Figure 4**b. It is shown that the frequency has a weak angular dependence for the [Co/Ni] at positive current. All frequencies exhibit the red-shift as the current increase which is from the negative nonlinearity N. [57] In contrast, at negative current, the dynamics of the exchange-spring layer shows an obvious angular dependence. The frequency decreases from 12.5 GHz to 5 GHz as the angle increases. All frequencies at each angle showed here exhibit red-shift behavior, indicating the negative nonlinearity of the exchange spring at the current angles and field. The extracted linewidth and power at the representative angle 50° are presented in **Figures 4**c and d, respectively. The linewidth can be as low as 10 MHz and up to 200 MHz. The power in **Figure 4**d increases as current for both [Co/Ni] and [Co/Pd]-Co layers. We have achieved the maximum power $P_{max} = 150$ pW from the exchange-spring layer, higher than the [Co/Ni] layer ($P_{max} = 10$ pW), which could be due to the larger exchange-spring precession angle than that of the [Co/Ni] layer. We also establish that the power for our bipolar STNOs is 2-3 orders of magnitude higher than that of bipolar spin-Hall nano-oscillators, while keeping the linewidth at the same level. [61] The power should be possible to optimize further by synchronizing multiple STNOs. [29,31,32,62]

## CONCLUSIONS

We have presented an experimental study of exchange-spring STNOs involving the [Co/Pd]-Co/Cu/[Co/Ni] spin valve. By using a [Co/Pd]-Co exchange spring as a reference layer, we observe both magnetic droplets and FMR-like oscillations in the [Co/Ni] free layer at both perpendicular and tilted angles, which are otherwise hard to detect in conventional all-PMA STNOs. Importantly, due to the strong exchange coupling field, the auto-oscillating frequency can reach as high as 10.8 GHz in a low-to-zero field regime. At t-AP states, this type of exchange-spring STNOs serves as bipolar oscillators operating at both positive and negative currents, with a microwave emission power up to 150 pW. With the benefits above, we believe this type of exchange-spring STNO greatly extends the design freedom of STNOs by selecting reference and free layers with highly different anisotropies, which is promising for applications in energy-efficient spintronics devices.

**METHODS**

**STNOs' fabrication.** The film stacks were deposited on thermally oxidized Si wafers by magnetron sputtering, consisting of a seed layer of Ta (4 nm)/Cu (14 nm)/Ta (4 nm)/Pd (2 nm), a free layer of [Ni (0.68 nm)/Co (0.22 nm)]×4, a spacer of Cu(7 nm), an exchange-spring reference layer of [Co (0.35 nm)/Pd (0.7 nm)]×5/Co(3 nm) and a cap layer of Cu(2 nm)/Pd(2 nm). We simplify this type of spin-valve multilayers to [Co/Pd]-Co/Cu/[Co/Ni], as presented in **Figure 1**. Using optical lithography and ion-milling etching techniques, 8×16 μm mesas were patterned on the stacked wafer and capped by a 30-nm-thick $SiO_2$ film as an insulator by chemical vapor deposition (CVD). Then, electron beam lithography (EBL) and the reactive ion etching (RIE) were used to fabricate nano-contacts through $SiO_2$ on top of each mesa, with a circular diameter varying from 50 to 150 nm. Finally, Cu(500 nm)/Au(100 nm) was deposited for the top electrode by using lift-off processing.

**Magnetic and electrical characterization.** Alternating Gradient Magnetometer (AGM) was used to measure the magnetization hysteresis loops of the unpatterned material stacks. The external fields were swept normal to the thin-film plane. Ferromagnetic resonance (FMR) was carried on the stacks by commercial NanOsc Phase FMR40. The frequency was swept from 2 to 40 GHz at fixed magnetic fields. dc and microwave measurements of the fabricated STNOs were carried out using our custom-built 40-GHz probe station. It allows the manipulation of magnetic field strength (up to 0.65 T), polarity, and direction. The device is connected through a ground–signal–ground (GSG) probe. The direct current, using a Keithley 6221 current source, was injected into the probe

(so as the device) through a 40-GHz bias-tee. The dc voltage was measured with a Keithley 2182 nano voltmeter. Here, we define the negative current as the electron flow from the free layer to the exchange-spring layer. When the current generates enough spin-transfer torque, auto-oscillation happens and microwave signals are emitted. These microwave signals were decoupled from the *dc* voltage via a bias-tee and then amplified using a low-noise amplifier before recorded by a spectrum analyzer (R&S FSU 20 Hz–67 GHz).

**ACKNOWLEDGEMENTS**

S. J. acknowledges the financial support from the Natural Science Foundation of China (No. 621044196) and Basic Research Programs of Taicang (Grant No. TC2021JC19). S. C. acknowledges support from the National Research Foundation of Korea (NRF) grant, funded by the Ministry of Science and ICT, Korea (No. 2020R1F1A1049642, 2022M3F3A2A03014536). PKJW acknowledges the financial support from the Natural Science Foundation of Shanxi Province, China (No. 2021JM-042). The authors thank the support from the Swedish Research Council (VR), the Swedish Foundation for Strategic Research (SSF), the Göran Gustafsson Foundation, and the Knut and Alice Wallenberg Foundation.